\documentclass[twocolumn,amsmath,floats,showpacs,nofootinbib]{revtex4}
\usepackage[usenames]{color}
\usepackage{graphicx}% Include figure files
\usepackage{epstopdf}
\usepackage{textcomp}
\usepackage{hyperref}
\usepackage{multirow}
\usepackage{tikz}
\usepackage{rotating}
\hypersetup{colorlinks=true}

\begin{document}

\title{Point-contact spectroscopy of tantalum, molybdenum, and tungsten}

\author{L. F. Rybal'chenko, I. K. Yanson,  N. L. Bobrov, and V. V. Fisun}
\affiliation{Physicotechnical Institute of Low Temperatures, Academy of Sciences of the Ukrainian SSR, Kharkov\\
Email address: bobrov@ilt.kharkov.ua}
\published {(\href{http://fntr.ilt.kharkov.ua/fnt/pdf/7/7-2/f07-0169r.pdf}{Fiz. Nizk. Temp.}, \textbf{7}, 169 (1981)); (Sov. J. Low Temp. Phys., \textbf{7}, 82 (1981)}
\date{\today}

\begin{abstract}Electron-phonon interaction in superconducting transition metals (Ta, Mo, W) is investigated by the method of point-contact spectroscopy. The spectra obtained are compared with model densities of the phonon states $F(\omega)$ computed from experiments on neutron scattering, and with the tunnel electron-phonon interaction function $g(\omega)$ for Ta. Close agreement between the appropriate spectra is obtained for Ta and Mo, while a significant discrepancy is observed for W, which is the appearance of a blurred maximum instead of the two expected separated peaks on the point-contact spectra. The absolute intensities of a number of point-contact spectra $G(eV)$ and electron-phonon coupling constants $\lambda$ are calculated. For all three metals the values obtained turned out to be considerably below those expected. The possible reasons for this disagreement are discussed.


\end{abstract}

\maketitle
\section{INTRODUCTION}
The method of point-contact (PC) spectroscopy has been applied successfully in the investigation of the electron-phonon interaction (EPI) in noble, alkali, and some other metals in recent years \cite{1,2}. The advantage of this method is the possibility of a direct determination of the so-called PC EPI function $G(eV)$ from the second derivative of the current-voltage characteristic \cite{3}

\begin{equation}
\label{eq__1}
G(eV)={-\frac{{{d}^{2}}I}{{{V}^{2}}}(eV)}\Big{/}{\frac{4\pi {{e}^{3}}}{\hbar}{{\Omega }_\text{eff}}N(0).}
\end{equation}

Here $N(0)$ is the density of the electron states, $\Omega_\text{eff}=d^3/3$ is the effective volume of phonon generation for the model of a hole of diameter $d$. The $G$ function is associated with the known EPI function $g(\omega)=\alpha^2(\omega)F(\omega)$ (Ref. \cite{4}):
\begin{equation}
\label{eq__2}
G(\omega)\approx \langle K_i\rangle g(\omega),
\end{equation}
where the form factor $\langle K_i\rangle$ in the free electron model, averaged over the Fermi surface, equals
\begin{equation}
\label{eq__3}
\langle K_1\rangle=\frac{1}{4};\  \langle K_2\rangle=\frac{1}{2};\  \langle K_3\rangle=\frac{2}{3}\frac{l_i}{L}
\end{equation}
for the model of a pure hole, a pure and impure channel of length $L$, respectively ($l_i$ is the impurity electron mean free path).

The application of this method can turn out to be quite promising for the development of new superconducting compounds and alloys with a high critical temperature since knowledge of the function $G(\omega)$ [and $g(\omega)$ in terms of it] permits computation of the superconducting parameters within the framework of the Eliashberg theory \cite{5}.
Refractory superconducting transition metals are of greatest interest as initial materials, which governs the expediency of their investigation.

Investigation of the EPI in vanadium by using point-contacts \cite{6} was reported on earlier. There were obtained PC spectra with sufficiently clear phonon features; however, the background level remained high, which did not permit reliable evaluations of the absolute intensities of the $G$ functions and the electron-phonon coupling constants $\lambda$. Moreover, in many cases currents through the oxidized and contaminated PC sections, that are of tunnel or semiconducting nature, appeared in parallel to currents through the relatively pure metallic constriction whose nonlinear energetic dependence reflects the EPI. These parasitic leakage currents reduced the intensity and distorted the form of the PC spectra. Their appearance was explained by the quite high chemical activity of $V$ (especially relative to the gaseous impurities O, N, H), which resulted in considerable solubility of these elements in $V$ and the formation of a large quantity of compounds, including the most inferior and the non-stoichiometric which have poor dielectric or metallic properties.

The purpose of the present paper is to investigate the PC spectra of refractory superconducting transition metals (Ta, Mo, W) with lower chemical activity than $V$, and to determine the absolute values of $G$ and $\lambda$.

The circumstance that solubility of the most active elements O, N, H in the materials under investigation is not high as compared with vanadium, and even tends to zero in some cases, contributed to some degree to obtaining high-quality EPI spectra with low background level and zero-bias anomalies by using point-contacts fabricated by chemical and electrolytical etching.

\section{METHOD OF THE EXPERIMENT}
The point-contact was produced between two pointed bulky electrodes which were squeezed together by means of an elastic element directly in liquid helium by mechanical or electrical means. The electrodes were fabricated by using mechanical treatment and were polished chemically or electrolytically in the concluding stage. The purity of the initial materials, estimated from the resistance ratio $R_{300}/R_{res}$ was 20, 1500, and 2000, respectively, for Ta, Mo, and W.

By using traditional modulation techniques in the current source mode, the dependence of the second harmonic of the modulating signal on the voltage $V_2(V)$ was recorded on a two-coordinate recorder, where the dependence is associated with the PC of the EPI function in the free electron model by the relationship
\begin{equation}
\label{eq__4}
G(eV)\approx\frac{3}{8}\ \frac{V_2(eV)}{\sqrt{2}eV_1^2}\ \frac{\hbar v_0}{d},
\end{equation}
where $v_0$ is the Fermi velocity, $cm/sec$, $V_1$, $V_2$ are the effective values of the first and second harmonic voltages, $V$, and $d$ is the contact diameter, $nm$. All the measurements were executed at a temperature of 4.2~$K$. A magnetic field was used to suppress the superconductivity of Ta.

It must be noted that a metallic nature of the conductivity corresponding to the pure metal was not achieved in all the PC. To obtain high quality EPI spectra, i.e., spectra with small background level and zero-bias anomalies and without quite definite parasitic currents, as a rule, required careful selection of the etch composition and mode.

\section{RESULTS AND THEIR DISCUSSION}

\begin{figure}[]
\includegraphics[width=8.5cm,angle=0]{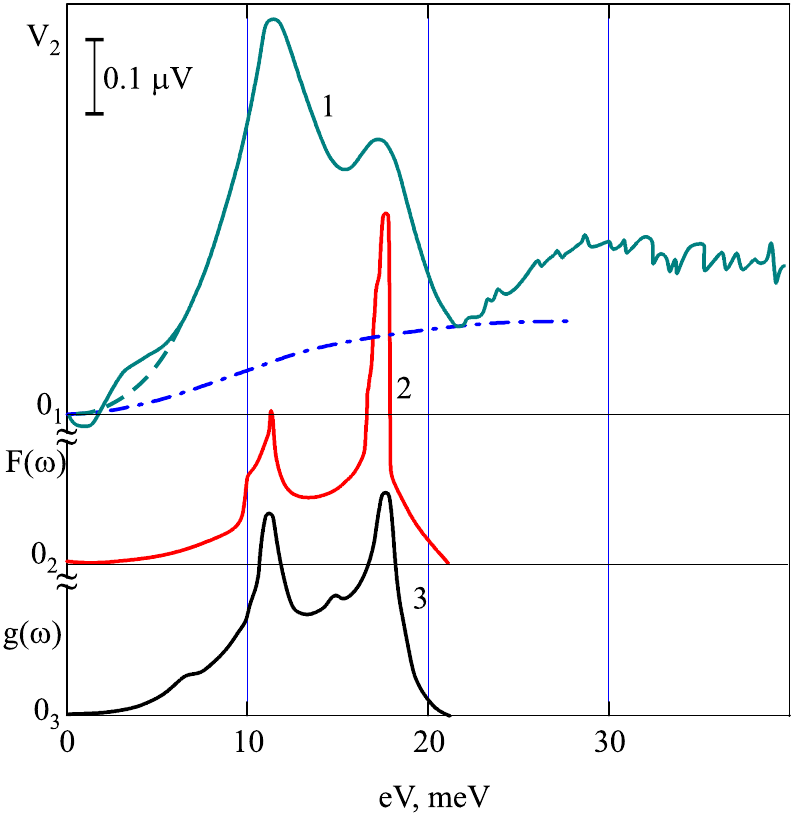}
\caption[]{Comparison of the Ta point-contact spectrum with the phonon density of states $F(\omega)$ and with the EPI tunnel function $g(\omega)$: 1) PC spectrum $V_2(eV)$, $R_0= 29~\Omega$, $V_1 = 0.81\ mV$, $H =18\ keV$, $T=4.2~K$; 2) $F(\omega)$; 3) $g(\omega)$
(Ref \cite{7}), $-\cdot -\cdot -\cdot-$ proposed dependence
of the background on the energy.}
\label{Fig1}
\end{figure}

Curve 1 in Fig. \ref{Fig1} is one of the typical PC spectra of Ta. It turned out to be possible to produce thin surface layers with good insulating properties in this metal, which permitted avoiding the passage of tunnel and semiconducting leakage current of noticeable magnitude through the PC. The MC with resistance $R_0$ from several to several tens of ohms were most optimal here. The PC spectra had a quite definite thermal nature without phonon singularities but with a significant background level on contacts with lower ohmic resistance.

The ratio of the intensities of the $L/T$ maxima corresponding to the longitudinal and transverse phonons, is close to the maximum achievable value for the spectrum presented in this paper $(\sim 0.6)$. This ratio turns out to be reduced considerably in certain spectra, and sometimes the magnitude of the longitudinal peak drops so much that just a small shoulder remains in its place. The position of the phonon singularities on the energy axis for different contacts can be considered fixed with an accuracy sufficiently high for PC spectroscopy; the $T$ peak at $11.5 + 0.5\ meV$, and the L peak at $17.5 + 0.5\ meV$. A blurred maximum, apparently corresponding to the so-called two-phonon processes, i.e., electron scattering with simultaneous generation of two phonons, was observed on the majority of spectra in the 30~$meV$ energy domain. Moderate zero-bias anomalies in the PC spectra of Ta usually were either of oscillating nature as in Fig. \ref{Fig1}, or consisted just of a negative section and did not exceed the anomalies presented in this figure, in magnitude.

The magnitude of the background $\gamma$ estimated as the ratio between the spectrum amplitude at energies rather greater than the two-phonon maximum and the magnitude of the greatest phonon peak (the $T$ peak in this case), was usually $\sim 0.3$, but had a considerably smaller value in individual cases. At this same time, in the majority of cases, the background did not saturate beyond the phonon spectrum but continued to grow according to an almost linear law.

Presented for comparison in Fig. \ref{Fig1} is also the phonon density of states $F(\omega)$ (curve 2), computed within the framework of the Born-Karman model from experiments on inelastic coherent neutron scattering, and the EPI tunnel function $g(\omega)$ (curve 3) calculated according to the Macmillan-Rowell program \cite{7}. Attention is turned to the quite close agreement between the positions of the phonon maxima on the energy axis for all three spectra. However, the ratio of their intensities remains different in each case.

Obtaining qualitative PC spectra (with low $\gamma$ and moderate zero-bias anomalies) was not fraught with substantial difficulties for Mo. As a rule, the parasitic currents had an insignificant magnitude, and the EPI spectra could be obtained at contacts with a broad band of resistances (from tenths to tens of ohms), although the most optimal were the contacts with several ohms resistance. A typical zero-bias anomaly on the PC spectra of Mo had a positive sign, where even in the case of its high intensity (for instance, if it exceeded the phonon maximum several times), distortion of the spectrum outside its limits would appear mainly in the shift of the phonon peaks towards high energies. The position of the $T$ peak at the same time moved from 21.5 to 24.5~$meV$, and of the $L$ peak, from 30 to 34~$meV$. For an electrical shorting by current pulses, the intensity of the positive zero-bias anomaly drops sharply, sometimes even with a sign change. The background does not ordinarily saturate in a PC with $R_0<1~\Omega$, but continues to grow as the voltage increases. It should also be noted that the two-phonon maximum was observed on the majority of PC spectra.

\begin{figure}[]
\includegraphics[width=8.5cm,angle=0]{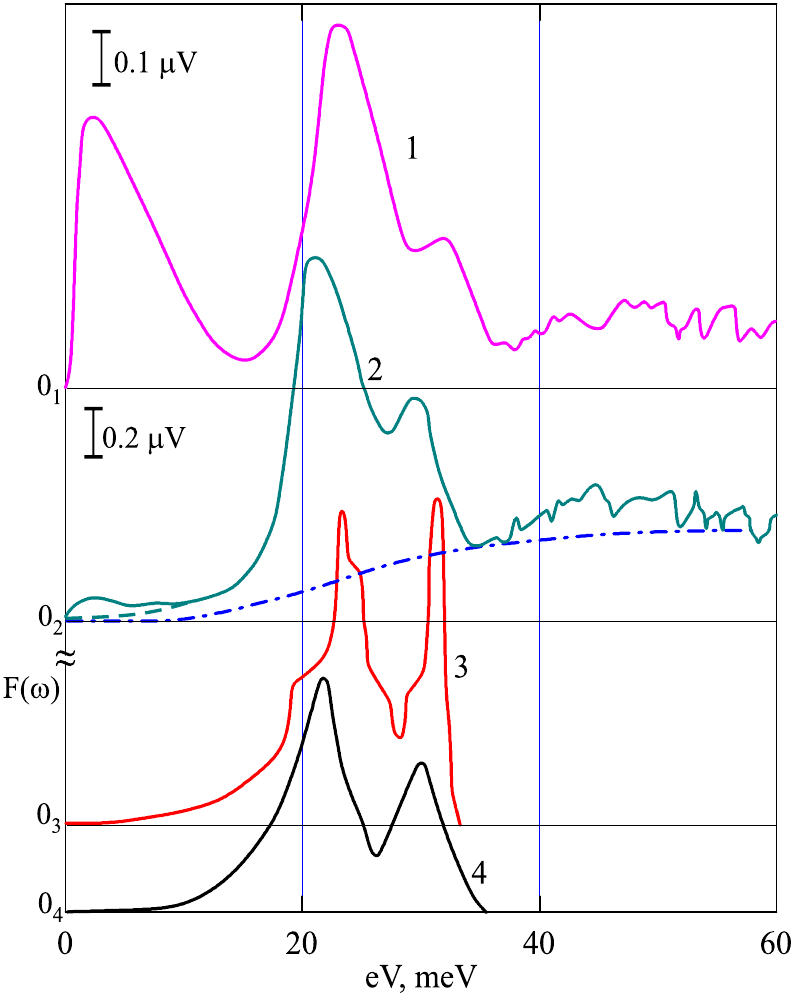}
\caption[]{ Comparison of Mo point-contact spectra with different zero-bias anomaly levels with two $F(\omega)$ computed within the framework of different models from neutron data: 1) $V_2(eV)$, $R_0 = 0.9\ \Omega$; 2) $V_2(eV)$, $R_0 = 5.5\ \Omega$, $V_1 = 0.63\ mV$, $T =4.2\ K$; 3) $F(\omega)$, obtained in the Bom-Karman model (Ref. \cite{8}); 4) $F(\omega)$ obtained in the angular forces model (Ref. \cite{9}.
}
\label{Fig2}
\end{figure}

Two Mo point-contact spectra with a different zero- bias anomaly level (curves 1 and 2) are presented in Fig.
\ref{Fig2}. Curves 3 and 4 are two functions $F(\omega)$ computed from neutron data within the framework of different models.
One of them (curve 3) was computed on the basis of the Bom-Harm an model \cite{8}, and the other (curve 4) within the framework of the angular forces model \cite{9} that takes into account also the noncentral interaction of ions. If it is assumed that the true PC spectrum is a spectrum with minimal zero-bias anomaly, then its closer agreement with $F(\omega)$ computed by the second model is detected. It must be emphasized that a diminution in $L$ peak intensity is also observed for the Mo point-contact spectra, although to a lesser degree than for Ta.

Tungsten and molybdenum are in the same subgroup of the periodic system of elements, and therefore, have similar physicochemical properties. It is hence completely legitimate that the properties of W and Mo point-contacts turn out to be similar in many respects. This is concerned mainly with the sign and magnitude of the zero-bias anomalies, the presence of parasitic current and the two-phonon maximum, as well as the behavior of the background. However, optimal values of $R_0$ for W turn out to be somewhat higher ($10-20\ \Omega$).

\begin{figure}[]
\includegraphics[width=8.5cm,angle=0]{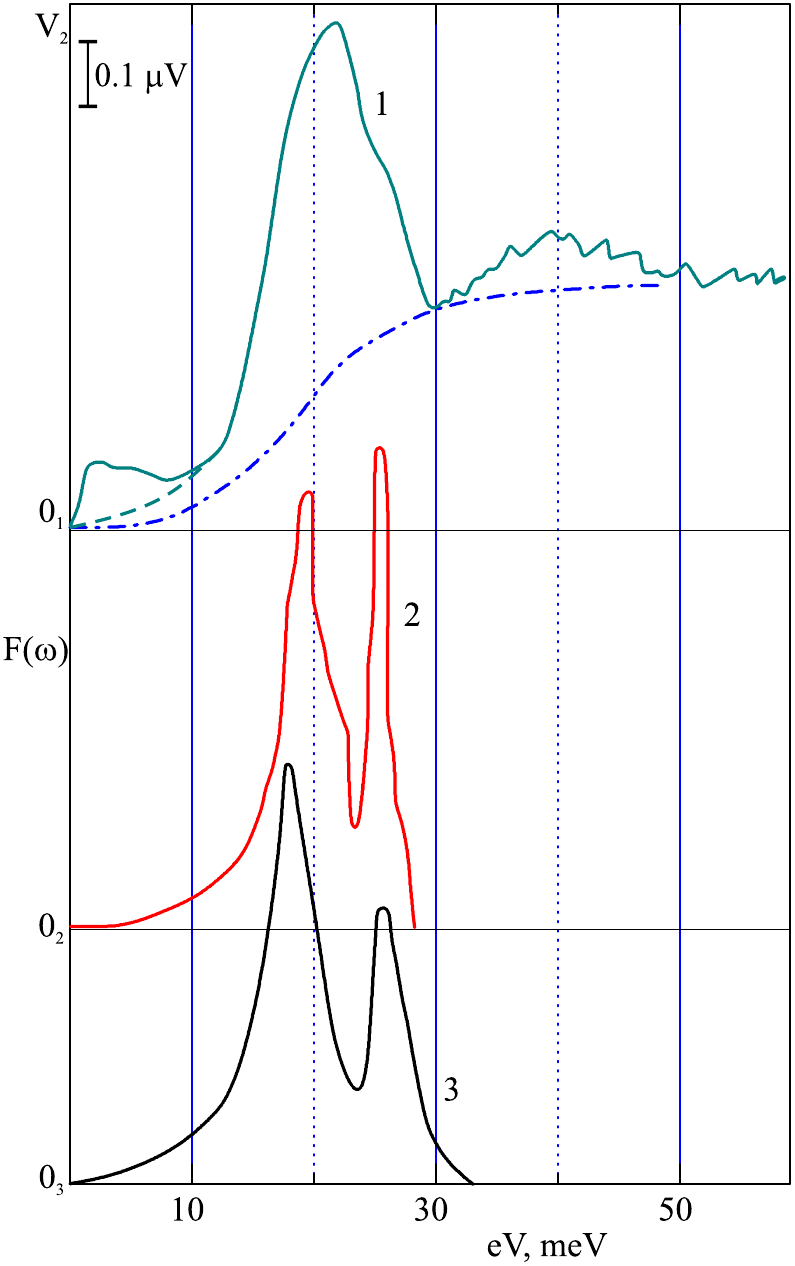}
\caption[]{Comparison of W point-contact spectrum with the functions $F(\omega)$ obtained by neutron spectroscopy: 1) $V_2(eV)$, $R_0 = 11.5\ \Omega$, $V_1 = 0.88\ mV$, $T = 4.2\ K$.; 2) $F(\omega)$ obtained with the Born-Karman model (Ref. \cite{10}); 3) $F(\omega)$ obtained with the angular forces model (Ref. \cite{9}; $-\cdot -\cdot -\cdot-$ proposed dependence of the background on the energy.
}
\label{Fig3}
\end{figure}
The distinguishing feature of the W point-contact spectra is the presence of one broad maximum (Fig. \ref{Fig3}, curve 1) instead of the two expected according to computations of the phonon density of states $F(\omega)$ from neutron data.
Curve 2 in Fig. \ref{Fig3} is the computed function $F(\omega)$ obtained in the Born-K arm an model \cite{10}, and curves 3 in the angular force model \cite{9}. Comparing these spectra shows that the $L$ and $T$ peaks of the $F$ functions appear on the PC spectrum only in the form of barely defined shoulders of the principal maximum, whose presence is not predicted by computation.

The absolute values of $G$ functions were calculated for a number of typical Ta, Mo, and W point-contact spectra by means of (\ref{eq__4}). Then multiplying the values of $G$ obtained by the form factor $\langle K_i\rangle$ we can obtain the function $g(\omega)$. Since the model of a circular hole corresponds most to our kind of contacts and we cannot estimate the degree of PC domain contamination, we shall provisionally consider the contact pure, i.e., we assume $\langle K \rangle =1/4$. The diameter of the contact in the model of a pure hole is
\begin{equation}
\label{eq__5}
d={{\left( \frac{16}{3\pi }\ \frac{\rho l}{{{R}_{0}}} \right)}^{1/2}}
\end{equation}
[the values $(\rho l)^{Ta} =3.5\cdot 10^{-12}\ \Omega\cdot cm^2$, $(\rho l)^{Mo} =11.7\cdot 10^{-12}\ \Omega\cdot cm^2$, $(\rho l)^W =21\cdot 10^{-12}\ \Omega\cdot cm^2$, $v_0^{Ta} = 0.24 \cdot10^8\ cm/sec$, $v_0^{Mo} = 0.76 \cdot 10^8\ cm/sec$,   $v_0^{W} = 0.76 \cdot 10^8\ cm/sec$ were used for the computations].

The section of the spectrum with zero-bias anomalies was replaced in the calculations by a curve dependent quadratically on the energy, which smoothly joined the remaining spectrum (these are the dashed lines in the figures). The foundation for such a procedure can be justified by the fact that precisely a quadratic dependence on the energy is observed for $eV > 0$ on PC spectra without zero-bias anomalies. Moreover, subtracted from the PC spectra is the background, for which thermal dependences $V_2(eV)$ without quite definite phonon singularities recorded repeatedly on the low-resistance contacts are taken (the dash-dot lines in the figures).

Now, having obtained, $g(\omega)$, the EPI constant can be evaluated
\begin{equation}
\label{eq__6}
\lambda =2\int\limits_{0}^{{{\omega }_{\max }}}{g(\omega )\frac{d\omega }{\omega }\ .}
\end{equation}
It turns out that the calculated values of $\lambda$ for each of the
three metals will fluctuate within broad limits, differing
by almost an order of magnitude for certain contacts,
hence the maximal values are $\lambda_{PC}^{Ta}\approx 0.065$,  $\lambda_{PC}^{Mo}\approx 0.12$, $\lambda_{PC}^{W}\approx 0.05$
These are considerably below the values obtained from other data (Ref. \cite{5}: $\lambda^{Ta}=0.65$,  $\lambda^{Mo}=0.41$, $\lambda^{W}=0.28$.

One of the fundamental reasons for obtaining diminished point-contact constants might be neglecting the contamination of the contact domain, whereupon the absolute intensities of the spectra, and therefore, $\lambda_{PC}$ as well are reduced $\sim l_i/d$ times. It hence becomes conceivable why the lowest value, compared to the expected values of $\lambda_{PC}$, was obtained for Ta since the initial Ta is the dirtiest of the three metals, and O, N, and H have the highest solubility limits therein. Moreover, the formation of compounds with a metal type of coupling (lower oxides, for example) through which the flow of currents similar to ohmic currents in nature can result in a reduction in $G$ and $\lambda$, is not excluded in the chemical or electrolytic etching of transition metals. Still another reason for such a reduction might be the imperfection in the computation method based on the free electron model that is poorly applicable to transition metals.

Since there is also a tunnel EPI function together with the model $F(\omega)$ for Ta, a deduction can definitely be made about the existence of significant suppression of the longitudinal phonon peak for its PC spectra. By the example of copper monocrystals it has been shown in Ref. \cite{4} that such suppression can occur at dirty or deformed contacts. An analogous result is contained in Ref. \cite{11}. in which the function $g(\omega)$ was studied for Nb in the superconducting state by the method of tunnel spectroscopy, and it was shown that the ratio of the $L$ and $T$ maxima diminishes for deformed specimens. Since there is no doubt about the dirtiness of Ta, it can turn out to be responsible for the diminution in $L/T$ also.

A reduction in the $L/T$ ratio is also observed for Mo point-contacts, but since there are no other spectra besides the computed $F(\omega)$ in this case, there is no possibility of judging confidently the degree of reduction in this ratio.

An additional maximum, whose appearance existing computational models cannot predict, apparently exists between the $T$ and $L$ peaks in the EPI spectrum of W.
\section{CONCLUSION}
The results in this paper show that PC spectroscopy of the superconducting transition metals Ta, Mo, and W in the normal state permits useful qualitative information to be obtained about the EPI at this stage of the investigation. To obtain quantitative parameters, methods must be developed to produce clean contacts or, at least, to control the degree of contamination of the contact region. Moreover, the method of computing $G$ and $\lambda$ should be corrected with the complexity of the transition metal band structure taken into account.
\section{NOTATION}
Here, $\omega$ is the phonon frequency; $F(\omega)$ is the phonon density of states; $g(\omega)$ is the electron-phonon interaction tunnel function; $G(\omega)$ is the point-contact electron-phonon interaction function; $\alpha^2(\omega)$ is the square of the electron-phonon interaction matrix element averaged over the Fermi surface; $\langle K_i\rangle$ is the form-factor averaged over the Fermi surface; $\lambda$ is the electron-phonon coupling constant; $d$ is the effective contact diameter; $l_i$ is the electron mean free path due to impurities; $v_0$ is the Fermi velocity; $V$ is the bias voltage across the contact; $V_1, V_2$ are the effective first and second harmonics of the voltage.

\end{document}